# Is Gravity an Entropic Force?

Shan Gao

Unit for HPS & Centre for Time, SOPHI, University of Sydney, Sydney, NSW 2006, Australia;
E-Mail: sgao7319@uni.sydney.edu.au

**Abstract:** The remarkable connections between gravity and thermodynamics seem to imply that gravity is not fundamental but emergent, and in particular, as Verlinde suggested, gravity is probably an entropic force. In this paper, we will argue that the idea of gravity as an entropic force is debatable. It is shown that there is no convincing analogy between gravity and entropic force in Verlinde's example. Neither holographic screen nor test particle satisfies all requirements for the existence of entropic force in a thermodynamics system. Furthermore, we show that the entropy increase of the screen is not caused by its statistical tendency to increase entropy as required by the existence of entropic force, but in fact caused by gravity. Therefore, Verlinde's argument for the entropic origin of gravity is problematic. In addition, we argue that the existence of a minimum size of spacetime, together with the Heisenberg uncertainty principle in quantum theory, may imply the fundamental existence of gravity as a geometric property of spacetime. This may provide a further support for the conclusion that gravity is not an entropic force.

**Keywords:** gravity; thermodynamics; entropic force; discrete spacetime

**PACS Codes:** 04.70.Dy; 04.20.Cv

## 1. Introduction

It is still a controversial issue whether gravity is fundamental or emergent. The solution of this problem may have important implications for a complete theory of quantum gravity. A remarkable indication for the nature of gravity comes from the deep study of black hole thermodynamics, which suggests that general connections between gravity and thermodynamics may exist [1–4]. Inspired by these theoretical developments, Jacobson argued that the Einstein equation can be derived from the proportionality of entropy and horizon area, together with the first law of thermodynamics, and he concluded that the Einstein equation is a thermodynamics equation of state [5]. Padmanabhan further showed that the equations of motion describing gravity in any diffeomorphism invariant theory can be given a thermodynamic re-interpretation, which is closely linked to the structure of action functional [6,7]. These results suggest that gravity may be explained as an emergent phenomenon and has a thermodynamics or entropic origin (see, e.g., [8] for a review). Recently Verlinde proposed a new argument for emergent gravity, mainly based on the holographic principle [9]. He argued and explicitly claimed that gravity is an entropic force, which is caused by a change in the amount of information associated with the positions of bodies of matter. This idea is interesting and, if right, may have important implications for the origin of gravity and its unification with the quantum. In this paper, we will critically examine the idea of gravity as an entropic force, focusing more on the physical explanation.

**2. Verlinde's Argument**

Verlinde's argument can be basically formulated as follows. Consider a small piece of a holographic screen. A particle of mass *m* approaches it from the side at which space has already emerged. First, it is assumed that before the particle merges with the microscopic degrees of freedom on the screen, it already influences the amount of information that is stored on the screen, and the corresponding change of entropy on the screen is:

$$\Delta S = 2\pi k_B \frac{mc}{\hbar} \Delta x \tag{1}$$

where $\Delta x$ is the displacement of the particle near the screen and comparable with the Compton wavelength of the particle, $k_B$ is the Boltzmann constant, *c* is the speed of light, and $\hbar$ is Planck's constant divided by $2\pi$. Next, it is assumed that the holographic principle holds, and the number of the used bits on the screen is:

$$N = \frac{Ac^3}{G\hbar} \tag{2}$$

where *A* is the area of the screen, and *G* is a constant that will be identified with Newton's constant later. Thirdly, it is assumed that the screen has a total energy *E*, which is divided evenly over the bits *N*, and the temperature of the screen *T* is determined by the equipartition rule:

$$E = \frac{1}{2} N k_B T \tag{3}$$

Lastly, it is assumed that the mass *M*, which would emerge in the part of space enclosed by the screen, satisfies the relativistic mass-energy relation:

$$E = Mc^2 \tag{4}$$

How does force arise then? Here Verlinde used an analogy with osmosis across a semi-permeable membrane. When a particle has an entropic reason to be on one side of the membrane and the membrane carries a temperature, it will experience an effective entropic force equal to:

$$F \Delta x = T \Delta S \tag{5}$$

Then he claimed that in the above interacting process between a holographic screen and a particle, the particle will also experience an entropic force in a similar way. Moreover, he showed that this force satisfies Newton's law of gravity, namely $F = G\frac{Mm}{R^2}$, based on Equations (1)–(5) and the area relation $A = 4\pi R^2$. Inspired by these interesting results, Verlinde thus concluded that gravity is an entropic force.

**3. Understanding Entropy Force**

In order to see whether gravity is an entropic force or not, we need to first understand the concept of an entropic force. An entropic force can be defined as an effective macroscopic force that originates in a thermodynamics system by its statistical tendency to increase entropy. As a typical example, let us see the elasticity of a polymer molecule, which was also discussed by Verlinde [9].

A polymer molecule is a large molecule composed of repeating structural units, typically connected by covalent chemical bonds. The simplest polymer architecture is a linear chain, and it can be modeled by joining together many monomers of fixed length, where each monomer can freely rotate around the points of attachment and direct itself in any spatial direction. Each of these configurations has the same energy. When the polymer molecule is immersed into a heat bath, it tends to put itself into a randomly coiled configuration since this configuration has higher entropy. There are many more such configurations when the molecule is short compared to when it is stretched into an extended configuration. The statistical tendency to reach a maximal entropy state then generates the elastic force of a polymer, which is a typical entropic force.

We can further determine the entropic force by introducing an external force $F$ to pull the polymer out of its equilibrium state and then examining the balance of forces. For example, one can fix one endpoint of the polymer at the origin and pull the other endpoint of the polymer apart along the $x$-axis. The entropy of the system can be written as $S(E,x) = k_B \log \Omega(E,x)$, where $\Omega(E,x)$ denotes the volume of the configuration space for the entire system as a function of the total energy $E$ of the heat bath and the average stretch length of the polymer (in this case the position $x$ of the pulled endpoint). One can then determine the entropic force by analyzing the micro-canonical ensemble given by $\Omega(E+Fx,x)$, where the external force is introduced as an external variable dual to the length $x$ of the polymer, and imposing the extremal condition for entropy. This gives $F = T\partial S/\partial x$, where the temperature $T$ is defined by $1/T = \partial S/\partial E$. By the balance of forces, the entropic force, which tries to restore the polymer to its equilibrium position, will be equal to the external force $F$. For the polymer the entropic force can be shown to obey Hooke's law, *i.e.*, the entropic force is proportional to the stretched length.

The entropic force can also be understood in terms of the first law of thermodynamics. The principle is an expression of energy conservation, according to which the increase in the internal energy of a system is equal to the amount of energy added by heating the system minus the amount lost as a result of the work done by the system on its surroundings. When the internal energy of a system does not change during the studied process, the law can be simply written as $F\Delta x = T\Delta S$. In the following we will analyze the entropic force of the polymer in terms of this equation.

We have three interacting systems in total, namely a heat bath, a polymer immersed in it, and an external system connected to the polymer. There is no entropic force when the polymer is in its equilibrium state with maximum entropy. Only when the polymer leaves its maximum entropy state can an entropic force appear. We assume that the internal energy of the polymer remains constant during the change of its entropy. When the polymer is pulled out of equilibrium by an external force, the work done by the force will be equal to the energy increase of the heat bath according to the first law of thermodynamics. In the l.h.s of the equation $F\Delta x = T\Delta S$, $F$ is the external force, $\Delta x$ is the displacement of the polymer, and $F\Delta x$ is the work done on the polymer by the external system. In the r.h.s of the equation, $T$ is the temperature of the polymer, $\Delta S$ is the entropy decrease of the polymer, and $T\Delta S$ is the heat loss of the polymer. The work is equal to the heat loss, and the internal energy of the polymer keeps constant. Moreover, the heat flows from the polymer to the heat bath, and the heat loss of the polymer is equal to the energy increase of the heat bath. As a result, the total energy of the three systems is conserved [10]. In short, energy flows from the external system to the polymer and further to the heat bath, and the corresponding causal chain is: External system → Polymer → Heat bath. During this process, the entropy of the polymer decreases, and the entropy of the heat bath increases. When keeping the polymer at a fixed length the entropic force will be equal to the external force, which is $F = T\Delta S / \Delta x$ according to the above equation, and their directions are opposite.

When one lets the stretched polymer gradually return to its equilibrium position, while allowing the force to perform work on the external system, the work done by the entropic force will be equal to the energy decrease of the heat bath according to the first law of thermodynamics. In the l.h.s of the equation $F\Delta x = T\Delta S$, $F$ is the entropic force, $\Delta x$ is the displacement of the polymer, and $F\Delta x$ is the work done by the polymer on the external system. In the r.h.s of the equation, $T$ is the temperature of the polymer, $\Delta S$ is the entropy increase of the polymer, and $T\Delta S$ is the heat of the polymer extracted from the heat bath. The work is equal to the heat gain, and the total energy of the three systems is also conserved. During this process, the entropy of the polymer increases, and the entropy of the heat bath decreases. Moreover, energy flows from the heat bath to the polymer and further to the external system, and the corresponding causal chain is Heat bath → Polymer → External system. By the equation $F\Delta x = T\Delta S$ we can also find the entropy force is $F = T\Delta S / \Delta x$.

Before ending this section we will stress several important features of entropic force, which are relevant to the analysis of Verlinde's argument that follows. First, entropic force results entirely from the statistical tendency of a thermodynamics system to increase its entropy, not from any energy effect. The energy of the system is conserved when the entropic force is in action. In this sense, entropic force has a purely entropic origin. But how can a force be generated without any energy effect? This leads us to the second important feature of entropic force, namely that the existence of a heat bath is indispensable for entropic force. Although an entropic force is independent of the details of the microscopic dynamics, its existence depends on the existence of interaction between the microscopic components of the studied system and environment or heat bath. For example, there is no entropic force for an isolated polymer in a vacuum. Due to the interaction, the system is constantly subject to random collisions from microscopic components of the heat bath, and each of these collisions sends the system from its current microscopic state to another. This will lead the system to its maximum entropy state, the state with the maximum number of microscopic states, and the statistical tendency to reach a maximum entropy state then generates an entropic force. In most familiar situations, the interaction has an eletromagnetic origin. In this sense, it may be not wholly right to say that entropic force has a purely entropic origin.

Thirdly, the magnitude of an entropic force is only related to the properties of the system and the surrounding heat bath, e.g., the size of the system and the temperature of the heat bath. For example, the entropic force of a polymer is proportional to its stretched length and the temperature of the heat bath. In particular, the magnitude of an entropic force is not related to the properties of any external system. For instance, the entropic force of a polymer is irrelevant to the mass of the external system connected to the polymer. Lastly, an entropic force always points in the direction of increasing entropy. An external force exerted on a system can point in the direction of decreasing its entropy, while the entropic force generated by the system must point in the direction of increasing its entropy. When the system reaches its maximum entropy, the entropic force becomes zero.

**4. Why Gravity Is Not An Entropy Force**

After we have understood what an entropic force is, we can next examine Verlinde's argument for the entropic origin of gravity. In his example of gravity, there are two systems, namely a holographic screen and a test particle. Verlinde argued that the gravitational interaction between the holographic screen and the particle is an entropic force. In order to see whether an entropic force exists in the example and whether gravity is an entropic force, we need to, parallel to the polymer example, answer the following two questions concerning Verlinde's example: Which is the heat bath? And which is the polymer?

Option 1: Verlinde's answer.

According to Verlinde, the holographic screen serves as the heat bath, and the particle can be regarded as the end point of the polymer that is gradually allowed to go back to its equilibrium position ([9], p. 25). He further suggested that the particle can be thought of as being immersed in the heat bath representing the screen in the holographic description, and by the time that the particle reaches the screen it will become part of the thermal state, just like the polymer.

As already admitted by Verlinde, however, it is not appropriate to view the screen as a heat bath. The reason is that the screen is not exactly at thermal equilibrium ([9], p. 26). If assuming the screen at an equipotential surface is in equilibrium, then the entropy needed to get the Unruh temperature will appear to be very high and violate the Bekenstein bound that states that a system contained in region with radius $R$ and total energy $E$ cannot have entropy larger than $ER$. Verlinde tried to solve this problem by rescaling the value of Planck's constant. This rescaling would affect the values of the entropy and the temperature in opposite directions: $T$ will get multiplied by a factor, and $S$ will be divided by the same factor. He also briefly proposed another possible solution based on a description that uses weighted average over many screens with different temperatures. Although Verlinde admitted that there is something to be understood, he still concluded that gravity is an entropic force.

In the following, we will point out some other problems of Verlinde's analogy between gravity and entropic force. First, although the particle can be regarded as a polymer in some sense, it seemingly has no well-defined temperature and entropy. As a result, it seems meaningless to talk about the entropy increase of the polymer-like particle or its statistical tendency to increase entropy, but the latter is required by the existence of an entropic force as we have seen in the polymer example. Furthermore, even if the particle has appropriate temperature and entropy, they will be very different from those of the screen, and thus the resulting entropic force for the polymer-like particle will be different from Newton's gravity. In fact, the temperature and entropy of the particle does not appear in Verlinde's derivation, where there are only the temperature and entropy of the screen. Secondly, during the process of the polymer-like particle going back to its equilibrium position ([9], p. 25), energy flows into the heat-bath-like screen. This is also inconsistent with the energy-flow feature of entropic force; energy should flow out of the heat bath for such a process. This leads us to the third problem. As we have known from the polymer example, during the process of a polymer returning to its equilibrium position, it is the entropy of the polymer that increases, while the entropy of the heat bath decreases. However, the entropy of the screen as a heat bath increases when the particle as a polymer returns to its equilibrium position. Therefore, there is also an obvious inconsistency in the analogy. To sum up, Verlinde's answer is problematic. It is improper to view the holographic screen as a heat bath and the particle a polymer. Moreover, no entropic force exists when assuming this view.

Option 2: The screen is taken as a heat bath, and the particle is taken as an external system.

Can an entropic force exist for this option? The answer is negative too. As we have seen from the analysis of Option 1, there is no convincing analogy between the screen and a heat bath. Thus, an environment suitable for the existence of entropic force is not available either for this option. Besides, the gravitational interaction is related to the mass of the external particle, while entropic force should only depend on the properties of the system (in this case the heat bath) and be irrelevant to the mass of any external particle. As a result, gravity cannot be an entropic force no matter whether entropic force exists for this option.

Option 3: The screen is taken as a polymer, and the particle is taken as an external system.

This option is suggested by the observation that the system whose entropy increases during the interacting process is the screen, not the particle, and it is the polymer, not the heat bath, that increases its entropy during similar process in the polymer example. Thus the screen is more like a polymer than like a heat bath. Besides, this option is also supported by the above observation that the gravitational interaction undergone by the particle points in direction of the entropy increase of the screen, which is consistent with the direction feature of entropic force. However, gravity cannot be an entropic force for this option either when considering the magnitude feature of entropic force. The gravitational interaction between the screen and the particle is related to the masses of both systems, while the entropic force originating from the polymer-like screen is irrelevant to the properties of the particle taken as an external system. In fact, there is no entropic force either for this option, as there is no heat bath here. As we have pointed out in the last section, a heat bath is indispensable for the existence of an entropic force.

Option 4: The screen and the particle are taken as the two ends of a polymer.

This option is consistent with the magnitude feature of an entropic force; the gravitational interaction between the screen and the particle is related to the properties of both systems, while the entropic force of a polymer also depends on the properties of the whole polymer, in this case including both the screen and the particle. However, there is an obvious objection to this option, namely that there is no thermal equilibrium between the screen and the particle. Moreover, as we have argued before, it is also problematic to view the particle as a polymer. Besides, since there is no heat bath, there is no entropic force either for this option. Note that although constant quantum vacuum fluctuations exist in the emergent space between the screen and the particle, the temperature of this Minkowski vacuum is precisely zero, and the in-between space is not a heat bath. Even if we assume the existence of an in-between gravitational field beforehand, and further regard it as an approximate heat bath, an environment suitable for the existence of entropic force is still unavailable. The reason is that the gravitational field is not in equilibrium with the screen and particle. Moreover, since the average temperature of the field is in general much lower than that of the screen or horizon, energy cannot spontaneously flow from the field to the screen due to a pure entropic reason.

After examining all these possible options, we find that there is no convincing analogy between gravity and entropic force in Verlinde's example. Neither the holographic screen nor the particle satisfies all the requirements for the existence of entropic force in a thermodynamics system.

Lastly, we will present a general objection to viewing gravity as an entropic force in Verlinde's example. The objection concerns the energy increase of the screen. During the interaction between the screen and the particle, the energy of the screen also increases along with its entropy increase [11].

Moreover, the amount of its entropy increase just corresponds to the amount of its energy increase. This already indicates that gravity cannot be a pure entropic effect, because, unlike the polymer example, energy is obviously input to the screen during the process, especially in the same amount as required by the increase of entropy. We can strengthen this conclusion by further analyzing the causal relationship between energy increase and entropy increase. It is well known that energy increase can cause entropy increase, but the (spontaneous) entropy increase cannot cause energy increase (on the contrary, energy will decrease if the entropy-increasing system does work. When no work is done by the entropy-increasing system, its energy is conserved in case of no input energy). Therefore, the entropy increase of the screen is in fact caused by its energy increase, not by a statistical tendency to increase entropy, which is essentially required for the existence of an entropic force. As a result, the interaction between the screen and the particle is not an entropic effect, and gravity is not an entropic force either [12].

The entropy increase of the screen results from its energy increase. Where does the increased energy come from then? When assuming a gravitational field exists between the screen and the particle as usual, it is easy to answer this question. It is the gravitational field that provides the increased (matter) energy for the screen [13]. In other words, the energy flow originates from the work done by the gravitational field through the force, gravity [14]. This is essentially different from the process of energy gain of the polymer. In the polymer example, the energy gain of the polymer comes from the heat bath through a pure thermodynamics process. Can a similar thermodynamics process explain the energy increase of the screen? The answer is negative. As we have argued in the analysis of Option 4, there is no well-defined heat bath here. Even if the gravitational field can be taken as an approximate heat bath, its average temperature will be in general much lower than that of the screen or horizon, and thus energy cannot spontaneously flow from the field to the screen due to a pure entropic reason. In a word, the increased energy of the screen can only come from the work done by the gravitational field.

Although the above argument seems reasonable, one question still needs to be answered before we can reach a definite conclusion, namely why the force $F$, derived from the formula $F\Delta x = T\Delta S$, is just gravity for the interacting process between a holographic screen and a particle when assuming the formula $\Delta S = 2\pi k_B \frac{mc}{\hbar} \Delta x$ as Verlinde did ([9], p. 7). In fact, this result can be readily understood after we have known that the entropy increase of the screen results from its energy increase and the energy comes from the work done by the gravitational field. Since $F = G\frac{Mm}{R^2}$ (Newton's law of gravity) and $T = \frac{1}{2\pi k_B}\frac{\hbar a}{c} = \frac{1}{2\pi k_B}\frac{\hbar GM}{cR^2}$ (Unruh's formula), there must exist the relation $\Delta S = 2\pi k_B \frac{mc}{\hbar}\Delta x$ in accordance with the formula $F\Delta x = T\Delta S$. Then it is not surprising that when assuming $\Delta S = 2\pi k_B \frac{mc}{\hbar}\Delta x$ (and Unruh's formula or holographic principle), Newton's law of gravity naturally follows [15].

It is worth noting that this argument does not depend on the distance between the screen and the particle, and the particle needs not to be near the screen. The general formula is $\Delta S = 2\pi k_B \frac{mcR^2}{\hbar R'^2}\Delta x$, where $R$ is the radius of the spherical screen, and $R'$ is the distance between the particle and the center of the screen [16]. Therefore, why Verlinde's "entropic force" is gravity is because it is just the work done by gravity that results in the increase in entropy [17].

To sum up, although Verlinde's derivation is right, it does not prove in physics that gravity is an entropic force [18]. Moreover, a detailed analysis shows that gravity is not an entropic force in the gravitational system he considered. In particular, the gravitational interaction between a holographic screen and a test particle is caused neither by the entropy increase of the screen nor by its statistical tendency to increase entropy; rather, the entropy increase of the screen is caused by gravity [19].

## 5. Further Discussions

The connections between gravity and thermodynamics seem so remarkable that one cannot help conjecturing that gravity has a thermodynamics or entropic origin. From a general point of view, however, this opinion is at least debatable. To begin with, the temperature and entropy of various horizons are all derived from the vacuum fluctuations of quantum fields *in* curved spacetime. Then how can one re-ascribe these emergent properties to the thermodynamics *of* spacetime itself? It seems that there is a huge gap between them in physics. Next, although the existing arguments based on thermodynamical analysis can derive the Einstein equation [20], they do it only with the help of the principle of equivalence together with some other assumptions. In other words, they can answer how matter curves spacetime only after assuming matter indeed curves spacetime. They do not explain why matter curves spacetime or why gravity is a curved spacetime phenomenon, which is the very nature of gravity according to general relativity. Only after one explains this particular nature, can one understand the origin of gravity and answer whether gravity is emergent or not. In the following, we will present a tentative answer. It is argued that the existence of a minimum size of spacetime, together with the Heisenberg uncertainty principle in quantum theory, might help to explain why matter curves spacetime.

According to the Heisenberg uncertainty principle in quantum mechanics we have:

$$\Delta x \geq \frac{\hbar}{2\Delta p} \qquad (6)$$

The momentum uncertainty of a particle, $\Delta p$, will result in the uncertainty of its position, $\Delta x$. This poses a limitation on the localization of a particle in the nonrelativistic domain. There is a more strict limitation on $\Delta x$ in the relativistic domain. A particle at rest can only be localized within a distance of the order of its reduced Compton wavelength, namely:

$$\Delta x \geq \frac{\hbar}{2m_0 c} \qquad (7)$$

where $m_0$ is the rest mass of the particle. The reason is that when the momentum uncertainty $\Delta p$ is greater than $2m_0 c$ the energy uncertainty $\Delta E$ will exceed $2m_0 c^2$, but this will create a particle anti-particle pair from the vacuum and make the position of the original particle invalid. It then follows that the minimum localization length of a particle at rest can only be the order of its reduced Compton wavelength. Using the Lorentz transformation, the minimum localization length of a particle moving with (average) velocity $v$ is:

$$\Delta x \geq \frac{\hbar}{2mc} \text{ or } \Delta x \geq \frac{\hbar c}{2E} \tag{8}$$

where $m = m_0/\sqrt{1-v^2/c^2}$ is the relativistic mass of the particle, and $E = mc^2$ is the total energy of the particle. This means that when the energy uncertainty of a particle is of the order of its (average) energy, it has the minimum localization length. Note that Equation (8) also holds true for particles with zero rest mass such as photons.

The above limitation is valid in continuous spacetime; when the energy and energy uncertainty of a particle both become arbitrarily large, its localization length $\Delta x$ can still be arbitrarily small. However, the existence of a minimum size of spacetime will demand that the localization of any particle should have a minimum value $L_U$, namely $\Delta x$ should satisfy the limiting relation:

$$\Delta x \geq L_U \tag{9}$$

In order to satisfy this relation, the r.h.s of Equation (8) should at least contain another term proportional to the (average) energy of the particle, namely in the first order of $E$ it should be:

$$\Delta x \geq \frac{\hbar c}{2E} + \frac{L_U^2 E}{2\hbar c} \tag{10}$$

This new inequality, which can be regarded as one form of the generalized uncertainty principle [21], can satisfy the limitation relation imposed by the discreteness of spacetime. It means that the localization length of a pointlike particle has a minimum value $L_U$.

How to understand the new term demanded by the discreteness of spacetime then? Obviously it indicates that the (average) energy of a particle increases the size of its localized state, and the increase is proportional to the energy. Since there is only one particle here, the increase of its localization length cannot result from any interaction between it and other particles such as electromagnetic interaction. Besides, since the increased part, which is proportional to the energy, is very distinct from the original quantum part, which is inverse proportional to the energy, it is a reasonable assumption that the increased localization length does not come from the quantum motion of the particle either. As a result, it seems that there is only one possibility left, namely that the (average) energy of the particle influences the geometry of its background spacetime and further results in the increase of its localization length. We can also give an estimate of the strength of this influence in terms of the new term $\frac{L_U^2 E}{2\hbar c}$. This term shows that the energy $E$ will lead to an length increase $\Delta L \approx \frac{L_U T_U E}{2\hbar}$. This further implies that the energy $E$ contained in a region with size $L$ will change the proper size of the region to:

$$L' \approx L + \frac{L_U T_U E}{2\hbar} \tag{11}$$

This means that a flat spacetime will be curved by the energy contained in it. When the energy is equal to zero or there are no particles, the background spacetime will not be changed. Since what changes spacetime here is the average energy, this relation between energy and proper size increase change is irrelevant to the quantum fluctuations.

The above analysis based on the quantum uncertainty principle and the discreteness of spacetime might provide a possible basis for the Einstein equivalence principle. It implies that gravity is essentially a geometric property of spacetime, which is determined by the energy density contained in that spacetime, not only for macroscopic objects but also for microscopic particles. Moreover, the Einstein gravitational constant can also be determined in terms of the minimum size of discrete spacetime [22]. The result is:

$$\kappa = 2\pi \frac{L_U T_U}{\hbar} \qquad (12)$$

Note that this formula itself seems to also suggest that gravity originates from the discreteness of spacetime (together with the quantum principle that requires $\hbar \neq 0$). In continuous spacetime where $T_U = 0$ and $L_U = 0$, we have $\kappa = 0$, and thus gravity does not exist. It should be stressed that the existence of a minimum size of spacetime has been widely argued and acknowledged as a model-independent result of the proper combination of quantum mechanics and general relativity (see, e.g., [23] for a review). The model-independence of the argument for the discreteness of spacetime strongly suggests that discreteness is a more fundamental feature of spacetime. Therefore, it seems appropriate to analyze the implication of spacetime discreteness for the origin of gravity as above.

Certainly, if spacetime itself is emergent, then gravity must be also emergent as it is essentially a curved spacetime phenomenon. But even so, they should have corresponding microscopic elements in the pre-spacetime theory. On the other hand, as we have argued above, gravity is probably fundamental in the emergent spacetime. The argument not only holds true for microscopic particles, but also may apply to the bits living on a holographic screen as well. This may provide a further support for the conclusion that gravity is not an entropic force.

**Acknowledgments**

I am very grateful to Sabine Hossenfelder, Christian Wüthrich, Dean Rickles and Huw Price for helpful discussions and suggestions. I am also grateful to the Special Issue Guest Editor Jacob D. Bekenstein and two anonymous reviewers for their constructive comments. This work was supported by the Postgraduate Scholarship in Quantum Foundations provided by the Unit for History and Philosophy of Science and Centre for Time (SOPHI) of the University of Sydney.

8. Padmanabhan, T. Thermodynamical aspects of gravity: New insights. *arXiv* **2009**, arXiv:0911.5004 [gr-qc].
9. Verlinde, E.P. On the origin of gravity and the laws of Newton. *arXiv* **2010**, arXiv:1001.0785 [hep-th].
10. For an infinite heat bath, the corresponding entropy increase of the heat bath will be equal to the entropy decrease of the polymer, and the total entropy will be also conserved. Thus the entropic force is conservative [9].
11. In the polymer example, the energy of the polymer keeps constant when the entropic force does work (the energy for the work comes from the surrounding heat bath). This dissimilarity reconfirms the conclusion that the screen is not like a polymer.
12. Interestingly, [25] showed that if gravity is an entropic force as Verlinde argued, then the Coulomb force should be also an entropic force. But it is well accepted that the Coulomb force is a fundamental interaction transferred by virtual photons. Besides, the directions of the entropic force and the Coulomb force are opposite for two charges with different signatures. This has been identified as the problem of negative electromagnetic temperature [25,26]. As we think, these apparent contradictions also suggest that the idea of gravity as an entropic force is probably wrong. In addition, this result can also be taken as a support for our conclusion that it is gravity (and the Coulomb force) that result in the entropy increase of the screen, not the contrary.
13. Note that the energy of the screen only depends on the mass inside it and is irrelevant to the mass of the external particle. Moreover, the entropy of the screen does not include the entropy of gravitational field, and it is only the entropy of matter.
14. As a result, the energy and entropy of the gravitational field correspondingly decrease during the process.
15. It has been shown that this consistency also has a mathematical origin, and it is a consequence of the specific properties of solutions to the Poisson equation [26].
16. Verlinde also discussed this large-distance situation and implicitly presented the right formula ([9], p.10).
17. Verlinde also admitted that why his equations come out is because the laws of Newton have been ingredients in the steps that lead to black hole thermodynamics and the holographic principle (see [9], p.9). However, as we have argued, his attempt to reverse this argument was not successful.
18. It has been recently claimed that Verlinde's idea is supported by a mathematical argument based on a discrete group theory [27]. As we think, although the theory might provide a possible mathematical formulation of the holographical principle, it does not necessarily entail that gravity is an entropic force in physics. Besides, it is far from clear whether this formulation can naturally lead to general relativity and quantum field theory as two proper approximations.
19. As we think, our physical analysis also applies to the similar arguments proposed by other authors (e.g., [5–8]). Like Verlinde, Jacobson did not explicitly state the causal relationship between energy flux and entropy change either, though his analysis was more rigorous than Verlinde's [5]. It seems that Jacobson assumed the right causal chain, *i.e.*, energy flux → entropy change, as he said "the entropy is proportional to the horizon area" and "the area increase of a portion of the horizon will be proportional to the energy flux across it". However, he also reached a similar conclusion that the Einstein equation is a thermodynamics equation of state [5].

20. This fact should not surprise us very much, as the thermodynamics of gravitational systems such as a black hole are just derived in terms of general relativity and quantum field theory.
21. The argument here might be regarded as a reverse application of the generalized uncertainty principle (see, e.g., [23,24]). But it should be stressed that the existing arguments for the principle are based on the analysis of measurement process, and their conclusion is that it is impossible to *measure* positions to better precision than a fundamental limit. On the other hand, in the above argument, the position uncertainty or localization length of a particle is objective and real, and the discreteness of spacetime requires that the objective length has a minimum value, which is independent of measurement.